# Millimeter Wave Doppler Sensor for Nondestructive Evaluation of Materials

S. Liao, S. Bakhtiari, T. Elmer, B. Lawrence, E. R. Koehl, N. Gopalsami, and A. Raptis
Argonne National Laboratory
9700 S. Cass Avenue, Lemont, IL, 60439
(630) 252-8982; fax (630) 252-3250; e-mail bakhtiari@anl.gov

## INTRODUCTION

Resonance modes are intrinsic characteristics of objects when excited at those frequencies. Probing the resonance signatures can reveal useful information about material composition, geometry, presence of defects, and other characteristics of the object under test. Vibration spectra can be measured remotely with high degree of sensitivity using a millimeter wave (mmW) Doppler sensor and a remote excitation source. This novel nondestructive evaluation (NDE) method can work in a non-contact manner as an alternative or complementary approach to conventional NDE methods such as those based on acoustic/ultrasonic and optical techniques. Millimeter wave vibrometry can be used for a wide range of civil and national security applications. Examples include detection of defects and degradation for diagnostics and prognostics of materials components and rapid standoff inspection of shielded/sealed containers for contraband. In this paper, we evaluate the performance of a compact mmW vibrometer developed at Argonne. Our 94 GHz I-Q Doppler sensor monitors the mechanical vibration signature of the object under interrogation that is induced by continuous wave excitation. For proof-of-principle demonstrations, the test objects were mechanically excited by an electronically controlled shaker using sinusoidal waves at various frequencies ranging from DC to 200 Hz. We will present a number of laboratory test results and will discuss the method's applicability to some practical NDE applications.

## Resonance Signatures

Vibration eigen-modes at certain natural resonant frequencies are unique characteristics of an object experiencing mechanical excitation [1]. These eigen-modes and natural resonant frequencies are determined by the equivalent inertia mass $M_{eff}$ and stiffness $k_{eff}$, obeying the following law of physics,

$$M_{eff}^m(z)\frac{d^2\Delta}{d^2z} + k_{eff}^m(z)\Delta = 0 \qquad (1)$$

where $z$ is the axial coordinate of the object, e.g., axis of a cylinder and $\Delta$ is the vibration amplitude of the eigen-modes. Different eigen-modes have different values of $M_{eff}^m$ and $k_{eff}^m$ for the $m^{th}$ azimuthal eigen-mode number, thus giving rise to different eigen-mode patterns and natural resonant frequencies. In order to solve Eq. (1), boundary conditions have to be specified, e.g., free standing, simply support or clamped. As an example, using the boundary conditions associated with a free standing, empty cylinder, the natural resonant frequencies can be obtained from Eq. (1) resembling a simple oscillator,

$$\omega_{m,n} = n\sqrt{k_{eff}^m / M_{eff}^m} \qquad (2)$$

where $n$ is the axial eigen-mode number. From Eq. (2), the resonance frequency depends on the effective mass $M_{eff}^m$ and the effective spring constant $k_{eff}^m$, both of which strongly depend on the geometry, material and boundary conditions of the object under evaluation. Through measurement of the resonance signatures, frequency shift and number of resonances can effectively reveal information about the attributes of the object.

## Millimeter Wave Doppler Sensor

Our prototype millimeter Wave (mmW) sensor works at 94 GHz [2], providing a unique combination of sensitivity and long range for detection of vibration signals [3]. Compared to microwave techniques [4]-[28], mmW range has the advantages of a) shorter wavelengths providing greater sensitivity to small displacements and b) higher spatial resolution obtainable with a reasonable aperture size. Compared to laser Doppler technique, the main advantages of the mmW frequency range for remote sensing [3] include a) penetration through many optically opaque dielectric materials, b) low atmospheric attenuation allowing long-range operation, c) low sensitivity to surface condition of the object (optically coarse reflecting surfaces), and d) ease of alignment.

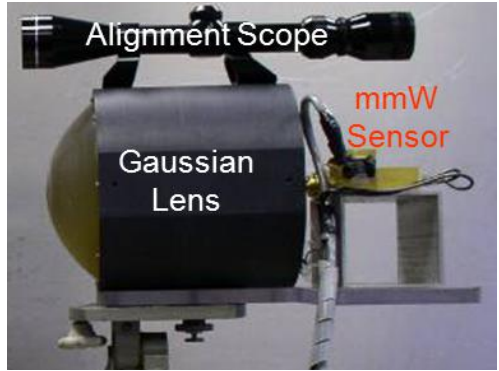

**Fig. 1: 94-GHz Doppler vibrometer with a Gaussian focus lens and an alignment scope.**

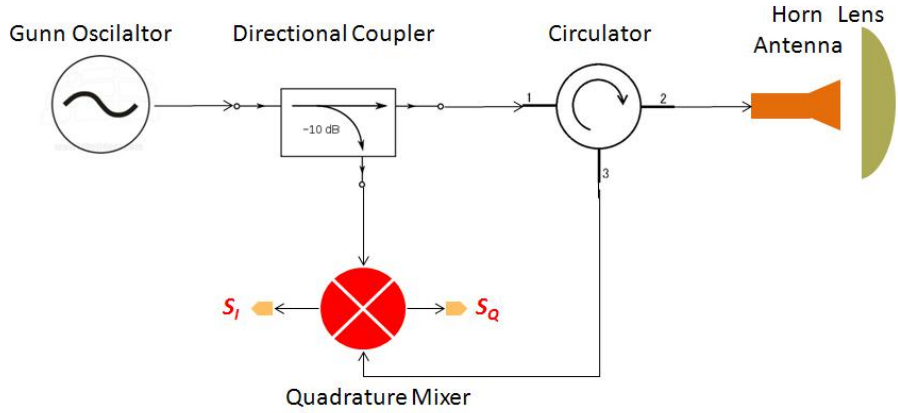

**Fig. 2: Functional block diagram of the 94-GHz Doppler vibrometer with a Gaussian optic lens antenna.**

A picture of the 94-GHz Doppler sensor system is shown in Fig. 1 and its functional block diagram is shown in Fig. 2. A W-band solid-state Gunn oscillator generates the reference and transmitted mmW signal. A fraction of the signal is fed to one of the inputs of the I-Q quadrature mixer. The remainder of the signal is fed to a circulator and corrugated horn antenna (2.39 mm in diameter) and is focused by a 6-inch-diameter dielectric lens on the target. The antenna system has a gain >25.0 dBi and a beam angle ~1degree. The reflected mmW signal is fed to the other input of the of the I-Q mixer, and is down-converted (homodyne) to both direct and quadrature channels, i.e., ($S_I$, $S_Q$). The phase $\varphi(t)$ can be expressed as [2]-[3],

$$S_I = A(t)\cos[\varphi(t)] \qquad (3)$$

$$S_Q = A(t)\sin[\varphi(t)] \qquad (4)$$

where $A(t)$ is the amplitude and $\varphi(t)$ is the associated phase. Combining Eq. (3) and Eq. (4) gives

$$\varphi(t) = \arctan\left[\frac{S_Q}{S_I}\right] \quad (5)$$

The signal or $\varphi(t)$ contains the object's Doppler frequency $f_d(t)$ information,

$$f_d(t) = \frac{1}{2\pi}\frac{d\varphi(t)}{dt} \quad (6)$$

## EXPERIMENTAL SETUP

The experimental setup for the mmW Doppler vibrometer is shown in Fig. 3. A picture of the laboratory components is shown in Fig. 4. A shaker is frequency-swept by a function generator from DC to 500 Hz, during a time period of $T = 999$ seconds. The frequency response of the object under evaluation is continuously monitored by the mmW system in real time. Both the voltage signal from the function generator and the I/Q signals ($S_I$, $S_Q$) from the mmW sensor are collected by LabVIEW™ software running on a Personal Computer (PC).

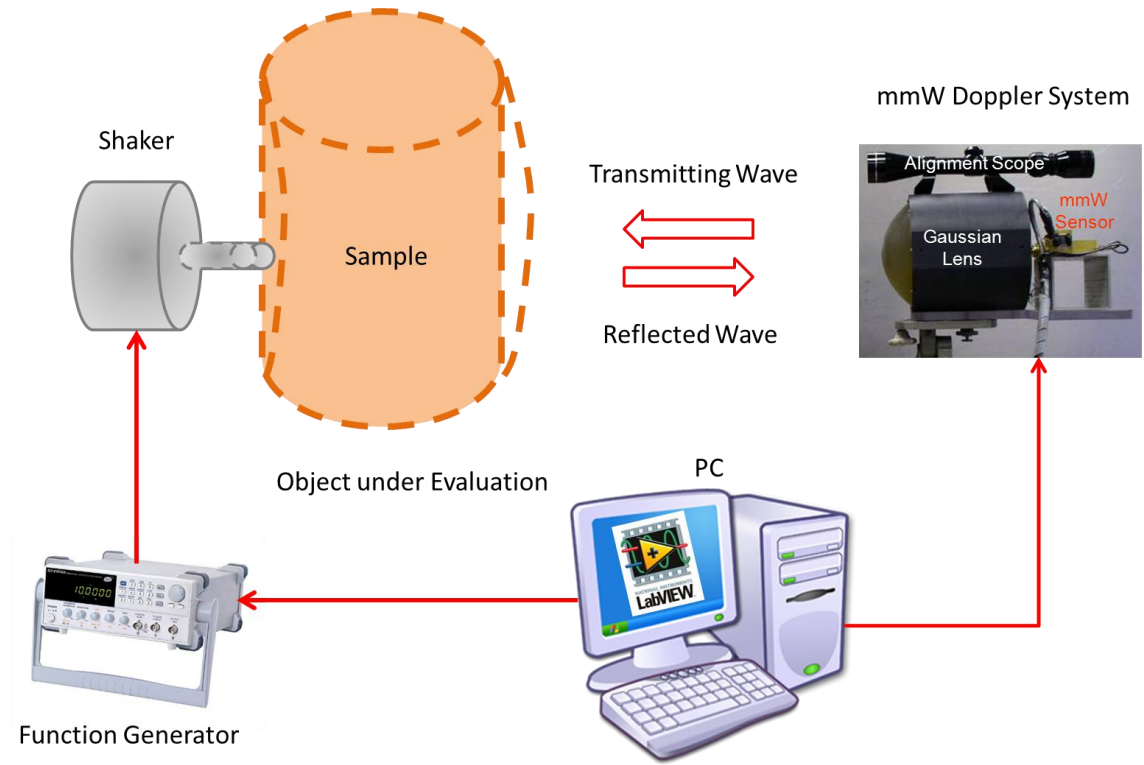

**Figure 3: Experimental setup for resonance signature measurement using a mmW Doppler sensor.**

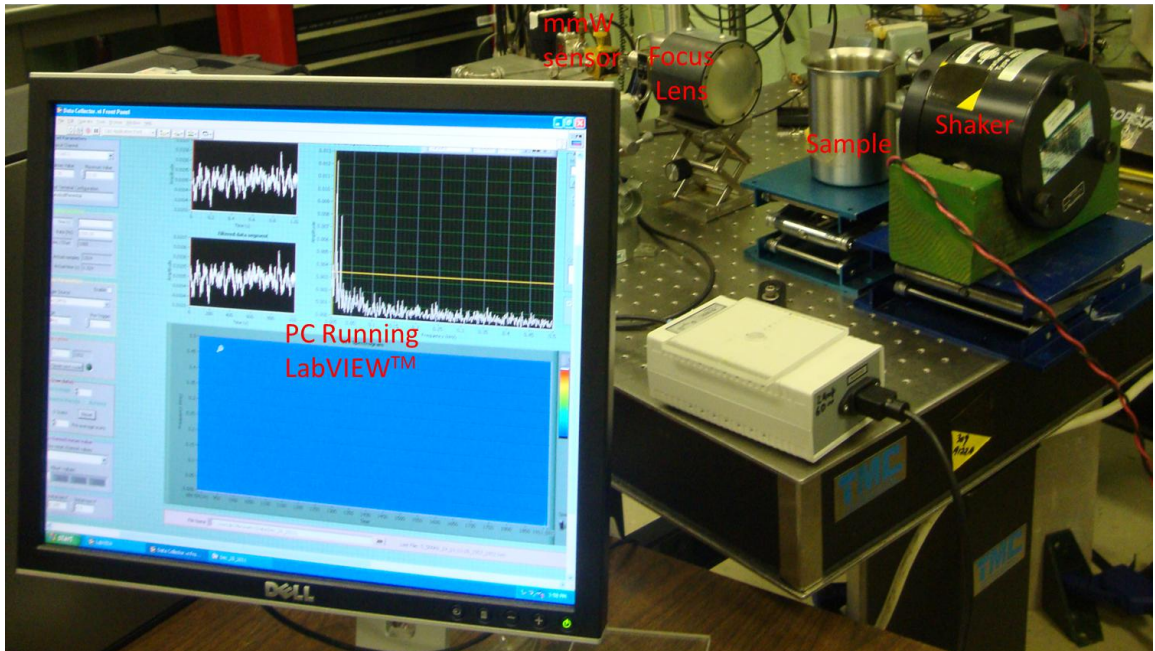
**Figure 4: Laboratory photo of the experimental setup shown in Fig. 3.**

## EXPERIMENTAL ASSESSMENTS
Two typical sets of experiments were carried out to assess the performance of the resonance signature NDE method; 1) simulated defects in plates made of different materials (plastic and aluminum), and 2) a cylindrical steel can loaded with different materials (water, and oil).

## Simulated Defects
Two types of manufactured defects have been investigated: circular and rectangular defects of different sizes on a) a 4-inch-by-4-inch aluminum plate, and b) a 4-inch-by-4-inch plastic plate. In all experiment, the plates were clamped on both sides as constraint support. Fig. 5 shows the resonance signatures for circular defects of case a): the resonant frequencies, as marked on each frequency spectra, are clearly different for the plate without (green) and with a ¼-inch diameter circular defect (blue), and with a ½-inch diameter circular defect (red). An example resonance frequency triplet from Fig. 5 is (61.52 Hz, 64.45 Hz, 66.41 Hz). Fig. 6 shows results for the rectangular defects for: the plate without defect (green), with 1/8-inch by ½-inch defect (blue), and with ¼-inch by ½-inch defect (red). An example resonance frequency triplet from Fig. 6 is (49.80 Hz, 42.97 Hz, 41.99 Hz). Similarly, Fig. 7 and Fig. 8 show the results for case b). Once again the resonant frequency shifts are clearly observable. Example resonance frequency triplets for Fig. 7 and Fig. 8 are (89.84 Hz, 88.87 Hz, 83.01 Hz) and (41.99 Hz, 38.09 Hz, 40.04 Hz) respectively.

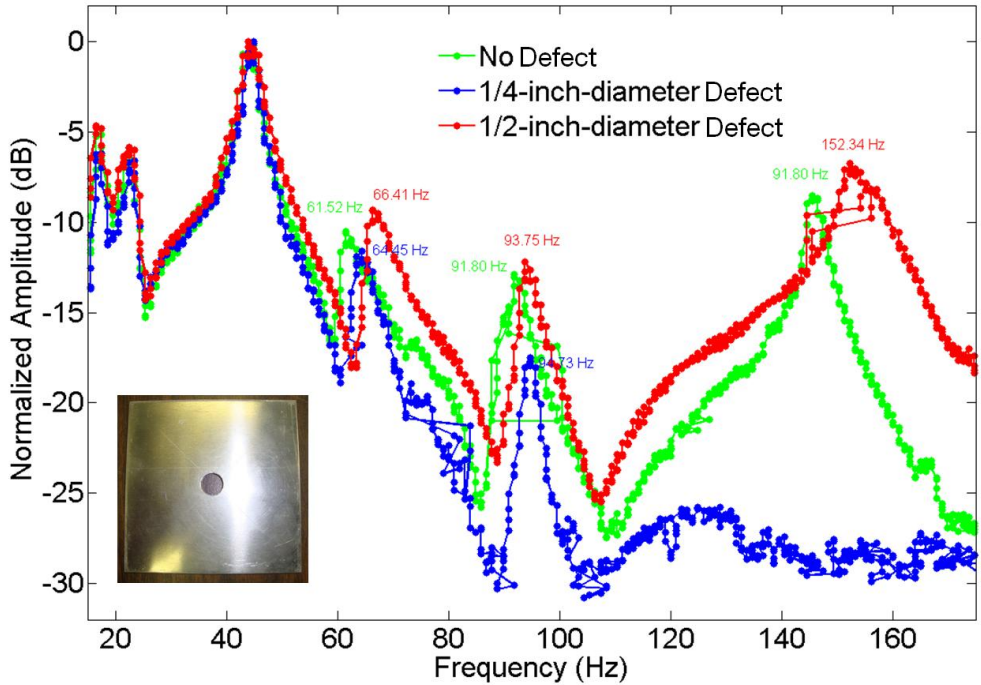

**Figure 5: Resonance signatures of a 4-inch-by-4-inch aluminum plate with manufactured circular defects. Also shown is the photo of a representative sample, i.e., a ½-inch circular defect.**

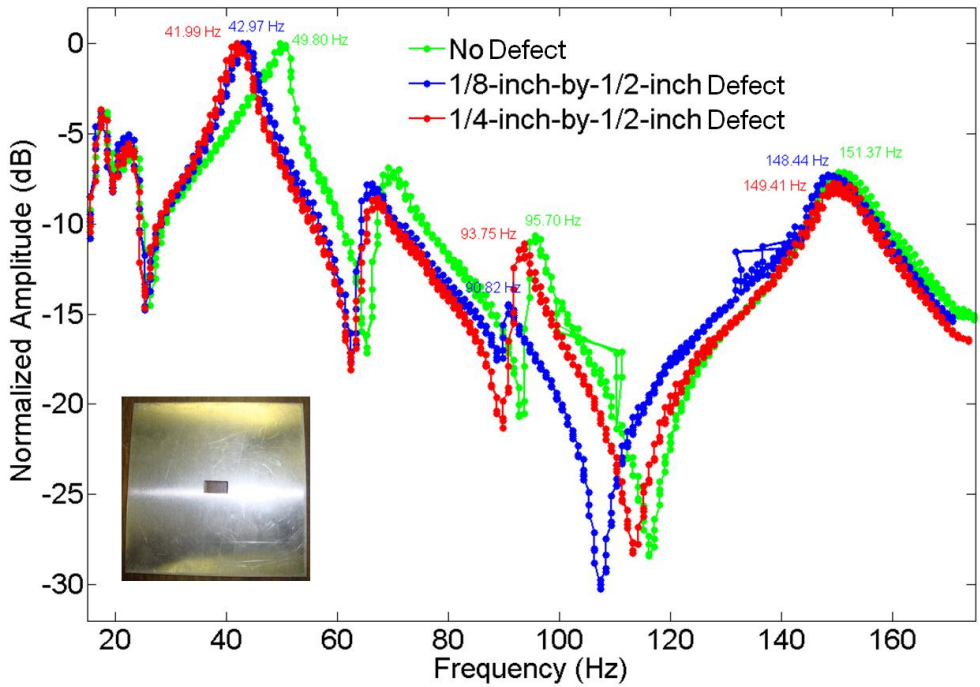

**Figure 6: Resonance signatures of a 4-inch-by-4-inch aluminum plate with manufactured rectangular defects. Also shown is the photo of a representative sample, i.e., a 1/4-inch-by-½-inch rectangular defect.**

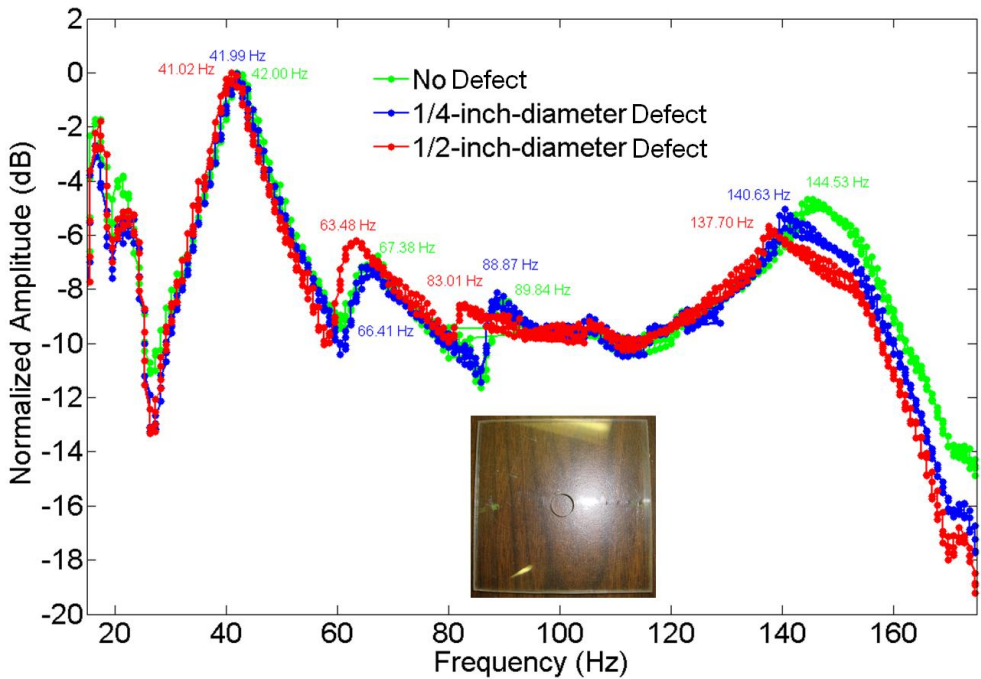

**Figure 7: Resonance signatures of a 4-inch-by-4-inch plastic plate with manufactured circular defects. Also shown is the photo of a representative sample, i.e., a ½-inch circular defect.**

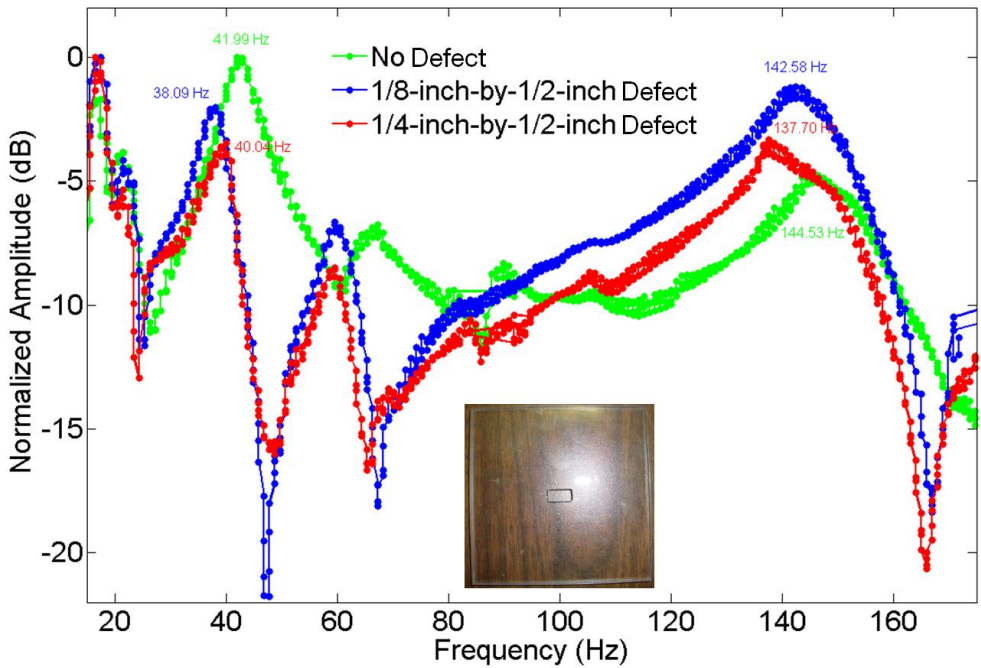

**Figure 8: Resonance signatures of a 4-inch-by-4-inch plastic plate with manufactured rectangular defects. Also shown is the photo of a representative sample, i.e., a 1/4-inch-by-½-inch rectangular defect.**

## Loaded Container

A 3-inch-diameter, 4.5-inch long cylindrical steel can was used in the experiment for two cases: a) fully-filled and half-filled with water; and b) fully-filled and half-filled with mineral oil. Fig. 9 shows the experimental results for case a), from which the resonant frequency shift is clearly detectable. For the case when the steel can is fully-filled with water, the resonant frequency is 65.43 Hz, compared to 72.27 Hz when it is half-filled with water. Fig. 10 shows similar results for case b). For the case when the steel can is fully-filled with oil, the resonant frequency is 57.62 Hz, compared to 73.24 Hz when half-filled with oil.

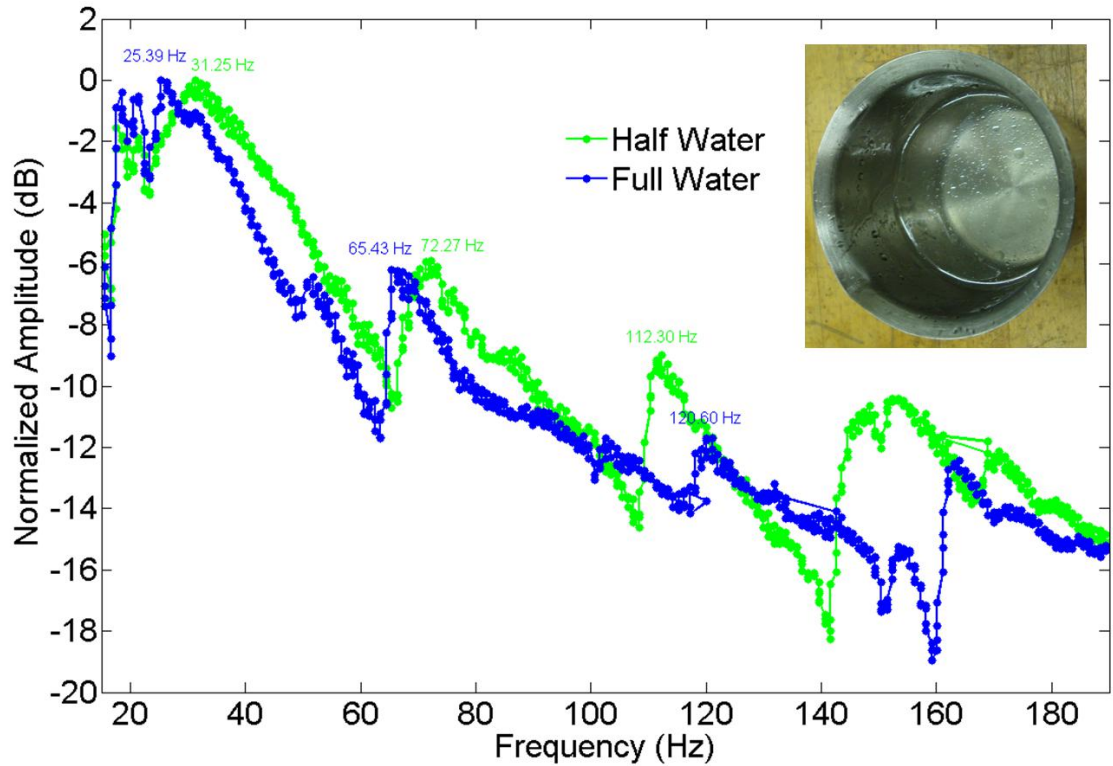

**Figure 9: Resonance signatures for a water-filled cylindrical steel can with a diameter of 3 inches and a height of 4.5 inches.**

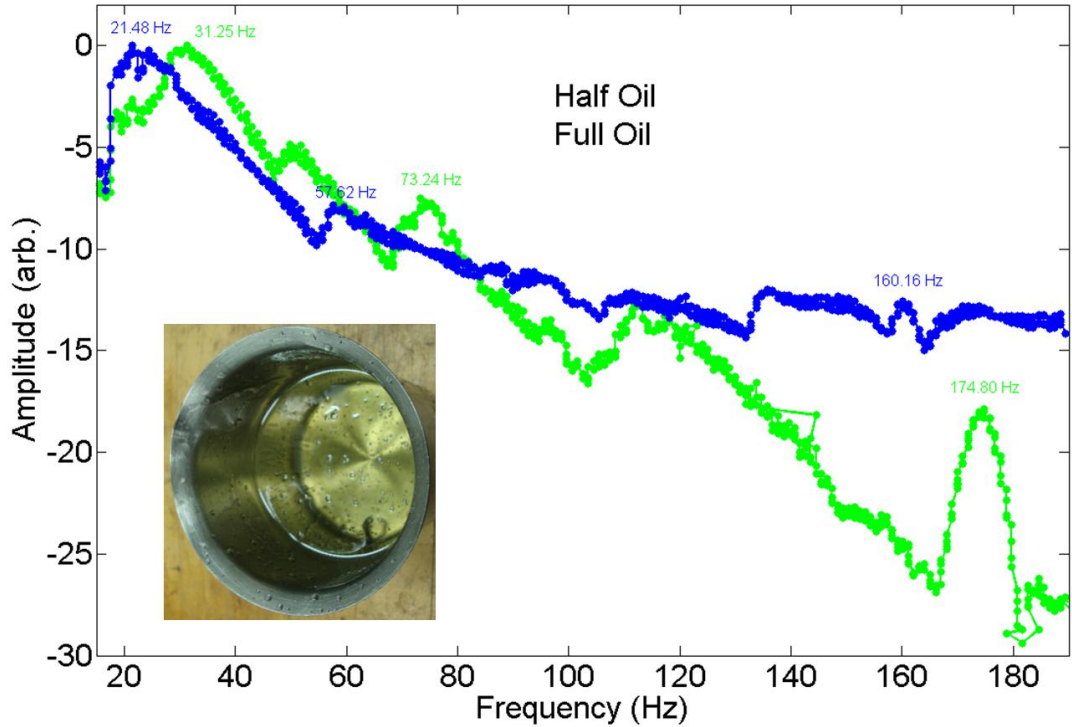

**Figure 10: Resonance signatures for an oil-filled cylindrical steel can with a diameter of 3 inches and a height of 4.5 inches.**

## CONCLUSION

A novel NDE method has been investigated for noncontact characterization of materials and components. The results suggest that the method provides a practical and simple approach for a wide range of standoff sensing applications. The measured signals are unique characteristic resonances of the object under evaluation. A 94-GHz Doppler sensor was used to monitor the resonance signatures in real time. Two sets of experiments have been performed to show the performance of the method for NDE applications. The experiments consisted of 1) Simulated defects of different sizes and shapes on plates made of different materials such as plastic and aluminum and 2) shielded/sealed materials (water and oil) inside a cylindrical steel can. The experimental results clearly show unique resonance signatures for all the objects tested in this work. The resonance frequency shifts of up to a few Hz were easily detectable by the sensor system. Based on the results of the investigations to date, it can be concluded that this NDE method can be used in many civil and national security applications such as detecting defects in multilayer structures and identification of the contents of shielded/sealed materials inside containers. Ongoing investigations in this work are associated with incorporation of an air-coupled excitation source such as high-power speaker or a laser, which will readily enable the method to be implemented as a standoff NDE tool.